\documentclass{article}

\usepackage{arxiv}

\usepackage[utf8]{inputenc} 
\usepackage[T1]{fontenc}    
\usepackage{hyperref}       
\usepackage{url}            
\usepackage{booktabs}       
\usepackage{amsfonts}       
\usepackage{nicefrac}       
\usepackage{microtype}      
\usepackage{lipsum}
\usepackage{graphicx}
\usepackage{newtxtext,newtxmath}
\usepackage{xcolor}
\usepackage{subcaption}
\usepackage[hang,flushmargin]{footmisc} 

\newcommand\blfootnote[1]{%
  \begingroup
  \renewcommand\thefootnote{}\footnote{#1}%
  \addtocounter{footnote}{-1}%
  \endgroup
}

\title{Mental state classification \\ using multi-graph features}

\author{
    Guodong~Chen$^{*, \#}$\\ Johns Hopkins University\\ gchen35@jhu.edu\\
    \And
    Hayden~S.~Helm$^{*, \dagger}$\\ Microsoft Research\\ v-haydenhelm@microsoft.com\\
    \And
    Kate Lytvynets \\
    Microsoft Research \\
    kalytv@microsoft.com \\
    \And
    \And
    Weiwei~Yang\\
    Microsoft Research \\  weiwya@microsoft.com\\
    \And
    Carey~E.~Priebe\\
    Johns Hopkins University \\ cep@jhu.edu\\
}

\begin{document}
\maketitle

\begin{abstract}
We consider the problem of extracting features from passive, multi-channel electroencephalogram (EEG) devices for downstream inference tasks related to high-level mental states such as stress and cognitive load. Our proposed method leverages recently developed multi-graph tools and applies them to the time series of graphs implied by the statistical dependence structure (e.g., correlation) amongst the multiple sensors. We compare the effectiveness of the proposed features to traditional band power-based features in the context of three classification experiments and find that the two feature sets offer complementary predictive information. We conclude by showing that the importance of particular channels and pairs of channels for classification when using the proposed features is neuroscientifically valid.

\end{abstract}


\section{Introduction}
\label{sec:intro}
Successful non-invasive Brain-Computer Interfaces (BCI) require solving a high dimensional, high frequency prediction problem. For EEG-based systems in particular, we have access to tens of streams of data that are  drastically attenuated by the skull. Hence, mining relevant predictive signal is a serious challenge. This is particularly true for passive tasks such as cognitive load and stress prediction where associated labels are not trial based and relatively    ``weak". 
\blfootnote{$^{*}$ denotes equal contribution \\
$^{\#}$ work partially done while working at Microsoft Research \\
$^{\dagger}$ corresponding author}

The most common approach to passive, non-invasive EEG-based BCI predictive tasks is to leverage neuroscientifically relevant features from the waveforms from each of the channels. For example, the alpha band (8-12 Hz) is known to be more active in stressed individuals and the theta (4-7 Hz) and low beta (13-20 Hz) bands are known to be active in fatigued persons. Simple functions of the relative masses of the sub-waveforms are also popular and useful features \cite{kamzanova2011eeg}. Similarly, functions of data from two channels of the EEG-device (such as frontal alpha asymmetry) have been shown to have different characteristics under different mental states and can be useful features for classification \cite{10.3389/fnbeh.2018.00166}. Both of these types of features rely on conventional neuroscience wisdom and are thus often hand-crafted for the particular predictive task.

Connectivity and correlation-adjacent features such as synchronization and EEG coherence between signals from different sensors have been explored in \cite{wei2007amplitude, wu2021resting}. However, the choice of different sensors to measure coupling strength is highly subjective and require neurophysiological a priori knowledge. Recent research has also investigated representing EEG signals as matrices \cite{congedo2017riemannian, yger2016riemannian}. These approaches derive the kernel on the Riemannian space using the average correlation matrix for a particular class. For other EEG-based tasks, such as classifying motor imagery \cite{oikonomou2017comparison}, techniques such as Common Spatial Filtering (CSP) and its variants \cite{koles1991quantitative,blankertz2007optimizing,ang2008filter} are used to learn a supervised projection from the set of sensors to an optimal subspace. Supervised methods embeddings such as CSP, at least empirically, often fail in passive BCI applications due to the supervisory signal being "weak" and not trial based.

On the flip side of conventional neuroscientific features is deep learning. Deep learning has achieved a state-of-art performance in fields such as speech recognition, visual object recognition, and object detection \cite{bengio2003out,lecun2015deep}. Many recent works have explored the use of convolutional neural network-based methods for automatic feature extractions in EEG-based BCIs \cite{lawhern2018eegnet,siddharth2019utilizing}. These approaches either use a pre-trained model such as VGG-16 \cite{simonyan2014very} to extract features from the power spectral density heat maps of different bands \cite{siddharth2019utilizing} or encapsulated an optimal spatial filtering in the network structure \cite{lawhern2018eegnet}. In our experience, as seems to be the experience of others \cite{lotte2018review}, current deep learning-based methods have not shown a significant improvement over methods utilizing conventional features in multiple BCI applications \cite{lotte2018review} and are notably less interpretable.

In this paper we propose a method that jointly learns a set of features from relationships between pairs of channels. The method leverages the temporal and spatial relationships between channels and will learn a different projection from the space of statistical-dependence matrices for each task -- yielding considerable flexibility compared to traditional hand-selected features. The process from going from EEG signal to feature vectors borrows heavily from recent developments in spectral-based multi-graph analysis \cite{Wang701755, gopalakrishnan2021multiscale}. We demonstrate in three different experiments each on two datasets that the proposed set of features can improve performance over standard band-based features and that the combination of the two nearly always improves over either individually. We argue that this improvement is due to the two sets of features capturing complementary predictive information. Lastly, we analyze the importance of each channel and each pair of channels and argue that the proposed feature set is neuroscientifically valid.

\subsection{Band-based features}
\label{subsec:BF}
In this subsection, we detail the traditional band frequency (BF) power spectral density approach. Consider a preprocessed multi-variate EEG time series $\{\boldsymbol{S}^{(t)} \in \mathbb{R}^{n_{c}}: t=1,2,\dots, T\}$, where $S_{j}^{(t)}$ represents signal from the $j$-th channel at time $ t $ and $ n_{c} $ is the number of channels. We first window the multi-variate EEG time series into potentially overlapping windows. 
Defining the window size to be $w$ and the overlap between windows to be $ h $, the EEG segment of the $k$-th window of the $j$-th electrode $ X_{j}^{(k)} \in \mathbb{R}^{w}$ is given by
\begin{equation*}
    X_{j}^{(k)} =  [S_{j}^{(k * (w-h) + 1)},S_{j}^{(k * (w - h) + 2)},\dots,S_{j}^{((k+1)* (w-h)+h)}].
\end{equation*} We let $ n_{w} $ be the number of windows after splitting the original times series in the time domain. For each channel $j$, we estimate the power spectral density $\{\hat{P}_{i}( X_{j}^{(k)}): k=1,2, \dots, n_{w}\}_{i=1}^{I}$ and sum the powers in $ I $  non-overlapping bands. We then normalize the band powers such that for each channel the features sum to 1,
\begin{equation*}
    \boldsymbol{F}_{ij}^{(k)}:=\hat{f}_{i}( X_{j}^{(k)}) =  \frac{\hat{P}_{i}( X_{j}^{(k)})}{\sum_{i=1}^{I} \hat{P}_{i}( X_{j}^{(k)})}.
\end{equation*}
Thus, for each time window $ k$, we have $ n_{c} \times I $ features. We consequently flatten this matrix into a single, $n_{c} \cdot I $ length vector $F^{(k)}$. Finally, because usually $ n_{c} \cdot I$-dimensional space will be relatively large given the amount of data we have access to, we project the feature vector into a low-dimensional space. The projection is learned via PCA of the available training data.  
\section{Methods}
\label{sec: methods}

\begin{figure}
    \centering
    \includegraphics[width=\linewidth]{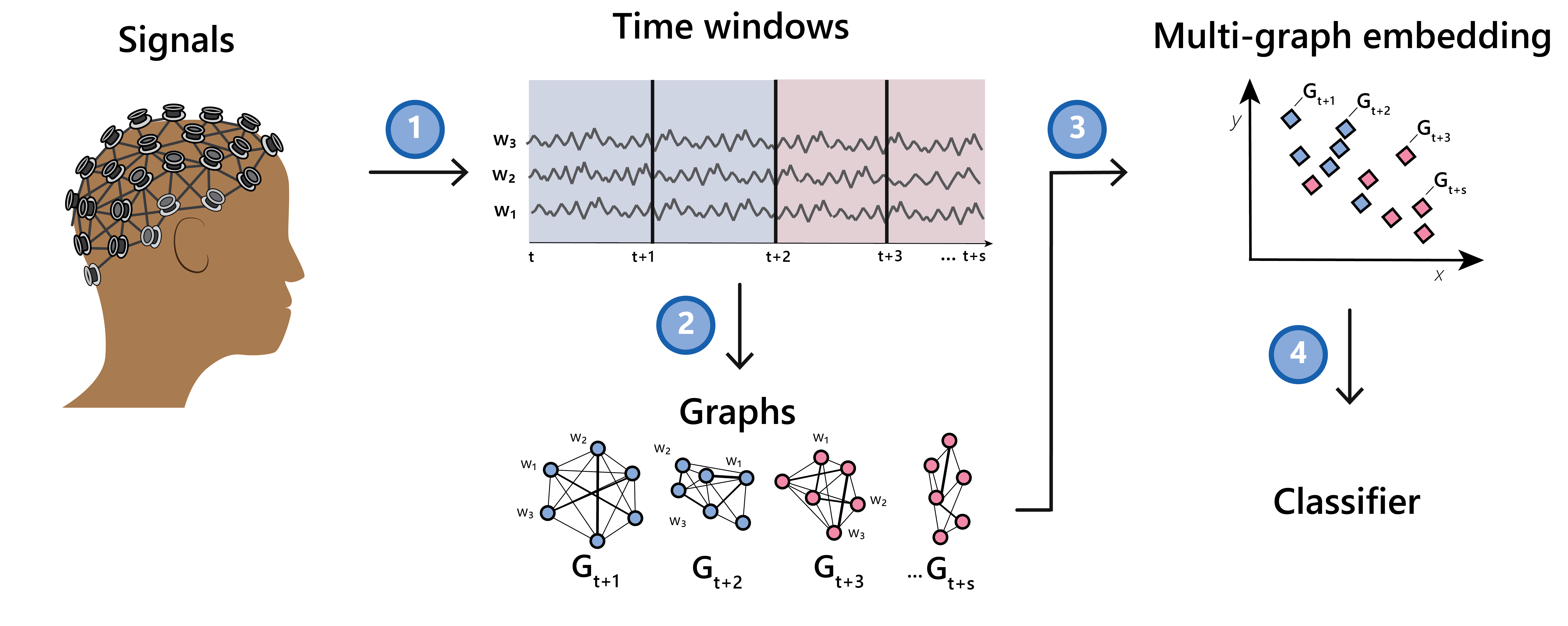}
    \caption{Illustration of going from a multi-channel EEG recording to a classifier via a time series of graphs.}
    \label{fig:method}
\end{figure}

The method we describe herein takes as input a collection of multi-channel time series and induces a network or graph on the set of channels. It then induces a graph on the set of networks and finally learns a single vector representation for each multi-channel time series. 
The representation can then be used as input to a classifier or other tools to aid downstream inference. See Figure \ref{fig:method} for an illustration of the method.

The described method is natively transductive and thus only learns a representation for the windows of the multi-channel time series that it has access to when learning the embedding. This can be limiting in applications where we want to apply the learned embedding from one session (or participant) to the data from another session (or participant). To alleviate this issue we describe an "out-of-sample" embedding that can take a previously unseen multi-channel time series and map it to the learned embedding space. We sometimes refer to the process of taking a windowed segment of EEG and projecting it into the appropriate embedding space as a feature mapping and the corresponding function as a feature map.

\subsection{Learning a representation from a time series of graphs}
\label{subsec:tsg}

We consider the windowed multi-variate time series described in Section \ref{subsec:BF}: 
\begin{equation*}
    X_{j}^{(k)} =  [S_{j}^{(k * (w-h) + 1)},S_{j}^{(k * (w - h) + 2)},\dots,S_{j}^{((k+1)* (w-h)+h)}].
\end{equation*}  

Let $ s: \mathbb{R}^{w} \times \mathbb{R}^{w} \to \mathbb{R} $ be a similarity function on objects in $ \mathbb{R}^{w} $ and $ A^{(k)} $ be the similarity matrix between pairs of the windowed times series for window index $ k $,
\begin{equation*}
    A^{(k)} = \begin{bmatrix} s(X_{1}^{(k)}, X_{1}^{(k)})  & \hdots & s(X_{1}^{(k)}, X_{n_{c}}^{(k)}) \\
    \vdots & \ddots & \vdots \\ s(X_{n_{c}}^{(k)}, X_{1}^{(k)}) & \hdots & s(X_{n_{c}}^{(k)}, X_{n_{c}}^{(k)}) \end{bmatrix}.
\end{equation*} The matrix $ A^{(k)} \in \mathbb{R}^{n_{c} \times n_{c}}$ can be thought of as a graph where the vertices are channels and the edge weight is the similarity score between the two channels for window $ k $.

Given the collection of graphs $ \{A^{(k)}\}_{k=1}^{n_{w}} $ there are numerous kernels that map a graph to a real-valued vector space \cite{kriege2020survey}. We use a spectral approach because it is relatively scalable and statistically principled in general settings \cite{JMLR:v18:17-448}. Hence, we induce a graph on the $ \{A^{(k)}\} $ by considering a pairwise similarity $ s': \mathbb{R}^{n_{c} \times n_{c}} \times \mathbb{R}^{n_{c} \times n_{c}} \to \mathbb{R} $ and corresponding pairwise similarity matrix
\begin{equation}
\label{eq: B}
    B = \begin{bmatrix} s'(A^{(1)}, A^{(1)})  & \hdots & s'(A^{(1)}, A^{(n_{w})}) \\
    \vdots & \ddots & \vdots \\ s'(A^{(n_{w})}, A^{(1)}) & \hdots & s'(A^{(n_{w})}, A^{(n_{w})}) \end{bmatrix}.
\end{equation}

Recall from results in statistical network analysis that when a graph $ B = U \Sigma V^{T} $ is a realization of a random dot product graph (RDPG) \cite{JMLR:v18:17-448} that the appropriately truncated left singular vectors of $ B $ scaled by the square root of the corresponding singular values are consistent estimates for the underlying latent positions $ Z \in \mathbb{R}^{n_{w} \times d} $  where $ Z^{(k)} \in \mathbb{R}^{d} $ is the latent position corresponding to window $ k $. In applications where the RDPG assumption is violated -- as is likely in ours -- reducing the dimensionality from $ n_{w} $ to dimension $ d < n_{w} $ in this way has shown to be useful for downstream inference tasks. 

We let $ \hat{Z} \in \mathbb{R}^{n_{w} \times d} $ be the estimate of the latent positions after embedding $ B $. These vectors will be the representations for the multi-channel time series that we use for inference. They will also serve as the basis representation for embedding previously unseen EEG windows. 

In our experiments we evaluate the effectiveness $ \hat{Z} $ as the representations used for mental state classification when letting $ s(\cdot, \cdot) $ be Pearson correlation and \begin{equation*}
    s'(A^{(k)}, A^{(k')}) = 1 - \frac{||A^{(k)} - A^{(k')}||_{2} - \min\{||A^{(k)} - A^{(k')}||_{2}\}_{k, k' \in \{1, .., n_{w}\}}}{\max\{||A^{(k)} - A^{(k')}||_{2}\}_{k, k' \in \{1, .., n_{w}\}} - \min\{||A^{(k)} - A^{(k')}||_{2}\}_{k, k' \in \{1, .., n_{w}\}}}. 
\end{equation*} We make some comments on potential improvements over these functions in Section \ref{sec:discussion}.

\subsection{Embedding out-of-sample windows}
\label{subsec:out-of-sample}

As mentioned above, the process to go from a windowed time series to a jointly learned vector representation in $ \mathbb{R}^{d} $ does not natively extend to previously unseen data. Though one could learn a new embedding every time new data is collected, this approach is relatively unscalable in both time and computation and of little use for real-time systems. To overcome this problem, ``out-of-sample" extensions of popular transductive embedding methods have been proposed to map new data to the space spanned by the original embeddings \cite{bengio2003out,trosset2008out,tang2013out,levin2018out}. We now describe the out-of-sample extension to the adjacency spectral embedding analyzed by Tang et al \cite{tang2013out}.

Given the original set of windowed time series $ \{X^{(k)}\}_{k=1}^{n_{w}} $, the learned embeddings $ \hat{Z} \in \mathbb{R}^{n_{w} \times d} $, a similarity function on data from the channels of the time series (i.e., $ s(\cdot, \cdot) $ from above), a similarity function on the graphs $ \mathbb{R}^{n_{c} \times n_{c}} $ (i.e., $ s'(\cdot, \cdot)) $, and a (potentially yet unseen) windowed time series $ X \in \mathbb{R}^{n_{c} \times w} $ with corresponding similarity matrix $ A $, we define the out-of-sample mapping to be
\begin{equation}
    T(X; \{X^{(k)}\}_{k=1}^{n_{w}}, \hat{Z}^{k}, s, s') := \sum_{k=1}^{n_{w}} s'(A, A^{(k)}) \left[(\hat{Z}^{T}\hat{Z})^{-1} \hat{Z}^{T}\right]_{k}.
\end{equation} As discussed in \cite{tang2013out}, in the context of graph embeddings the feature mapping $ T $ is considerably more time and computationally efficient than recalculating the singular value decomposition of the matrix defined in \eqref{eq: B} at the cost of a small performance degradation on downstream inference. Hence, for real-time applications $ T $ is preferred to recalculation.

\section{Results}
\label{sec:experiments}
We evaluate three sets of features -- band-power based features (``BF") of Section \ref{subsec:BF}, time series of graphs-based features (``TSG") of Section \ref{sec: methods}, and the concatenation of the two (``TSG + BF") -- in three different settings (in-session classification, non-constant querying, and transfer) in the context of two datasets (``Mental Math" and ``MATB-II"). Our analysis demonstrates the value of including both band-based and graph-based features for high level mental state prediction tasks across multiple experimental settings. After discussing the classification experiments we argue that the graph-based features are neuroscientifically reasonable via a channel and pair of channels importance study.

We note that though we interpret the results below as if the labels used to train the classifiers are only related to stress and cognitive load, there is a possibility of non-trivial confounding variables that could artificially improve performance on the intended task. We do not pursue a formal analysis of these confounding variables herein.

\subsection{Datasets}
\label{sec:datasets}

Mental Math \cite{data4010014} and MATB-II (proprietary) are representative of stress prediction and cognitive load prediction tasks, respectively, and are thus helpful in understanding the high-level mental state prediction capabilities of the three sets of features.  

\subsubsection{Mental Math}
\label{subsubsec:mm}
In the Mental Math study there are two recordings for each participant -- one corresponding to a resting state and one corresponding to a stressed state. For the resting state, participants counted mentally (i.e., without speaking or moving their fingers) with their eyes closed for three minutes. For the stressful state, participants were given a four digit number (e.g., 1253) and a two digit number (e.g., 43) and asked to recursively subtract the two digit number from the four digit number for 4 minutes. This type of mental arithmetic is known to induce stress \cite{noto2005relationship}.

There were initially 66 participants (47 women and 19 men) of matched age in the study. 30 of the participants were excluded from the released data due to poor EEG quality.
Thus we consider the provided set of 36 participants analyzed by Zyma et al \cite{zyma2019electroencephalograms}. The released EEG data was preprocessed via a high-pass filter and a power line notch filter (50 Hz). Artifacts such as eye movements and muscle tension were removed via ICA. 

For our analysis we further apply high pass (0.5 Hz) and low pass (30 Hz) filters and window the data into two and a half second chunks with no overlap. In our experiments we consider the two-class classification task \{stressed, not stressed\} and note that some of the windowed data in the training will be adjacent to some of the windowed data in the test set by virtue of how the experiment was set up. 

The TSG features that we use in the Mental Math experiments are based on the pairwise Frobenius norms between correlation matrices as described in Section \ref{sec: methods}. The number of components kept in the singular vectors is data-dependent and is the second ``elbow" of the scree plot of singular values estimated by Zhu and Ghodsi's method \cite{zhu2006automatic}.

The BF features we use are derived from the low (4.1 to 5.8 Hz) and high (5.9 to 7.4 Hz) theta bands; low (7.4 to 8.9 Hz), middle (9.0 to 11.0 Hz), and high (11.1 to 12.9 Hz) alpha bands; and low  (13.0 to 19.9 Hz), medium (20.0 to 25.0 Hz), and high (25.0 to 30.0 Hz) beta bands. Once the power of each of these bands is normalized by channel there are 19 (8) = 152 features per window that we then project into a lower dimension via PCA. As with the TSG features, we select the number of components via Zhu and Ghodsi's method. In our experiments this translates to between a 3 and 8 dimensional feature space that we then use for classification.

\subsubsection{MATB-II}

We also consider data collected under NASA's Multi-Attribute Task Battery II (MATB-II) protocol. MATB-II is used to understand a pilot's ability to perform under varying cognitive load requirements \cite{santiago2011multi} by attempting to induce four different levels of cognitive load -- no (passive), low, medium, and high -- that are a function of how many tasks the participant must actively tend to. 

The data we consider includes 50 healthy subjects with normal or corrected-to-normal vision. There were 29 female and 21 male participants and each participant was between the ages of 18 and 39 (mean 25.9, std 5.4 years). Each participant was familiarized with MATB-II and then actively underwent for two sessions containing three segments. The three segments were further divided into blocks with the four different levels of cognitive requirements. The sessions lasted approximately 50 minutes and were separated by a 10 minute break. 

The EEG data was recorded using a 24-channel Smarting MOBI device and was preprocessed using high pass (0.5 Hz) and low pass (30 Hz) filters and segmented in ten second, non-overlapping windows. In our analysis we consider the two-class problem \{no \& low cognitive load, medium \& high cognitive load\} and use the same feature extraction methods as described in Section \ref{subsubsec:mm}.

\begin{figure}[t!]
    \captionsetup[subfigure]{justification=centering}     
    \begin{subfigure}{0.49\textwidth}
     \includegraphics[width=\linewidth]{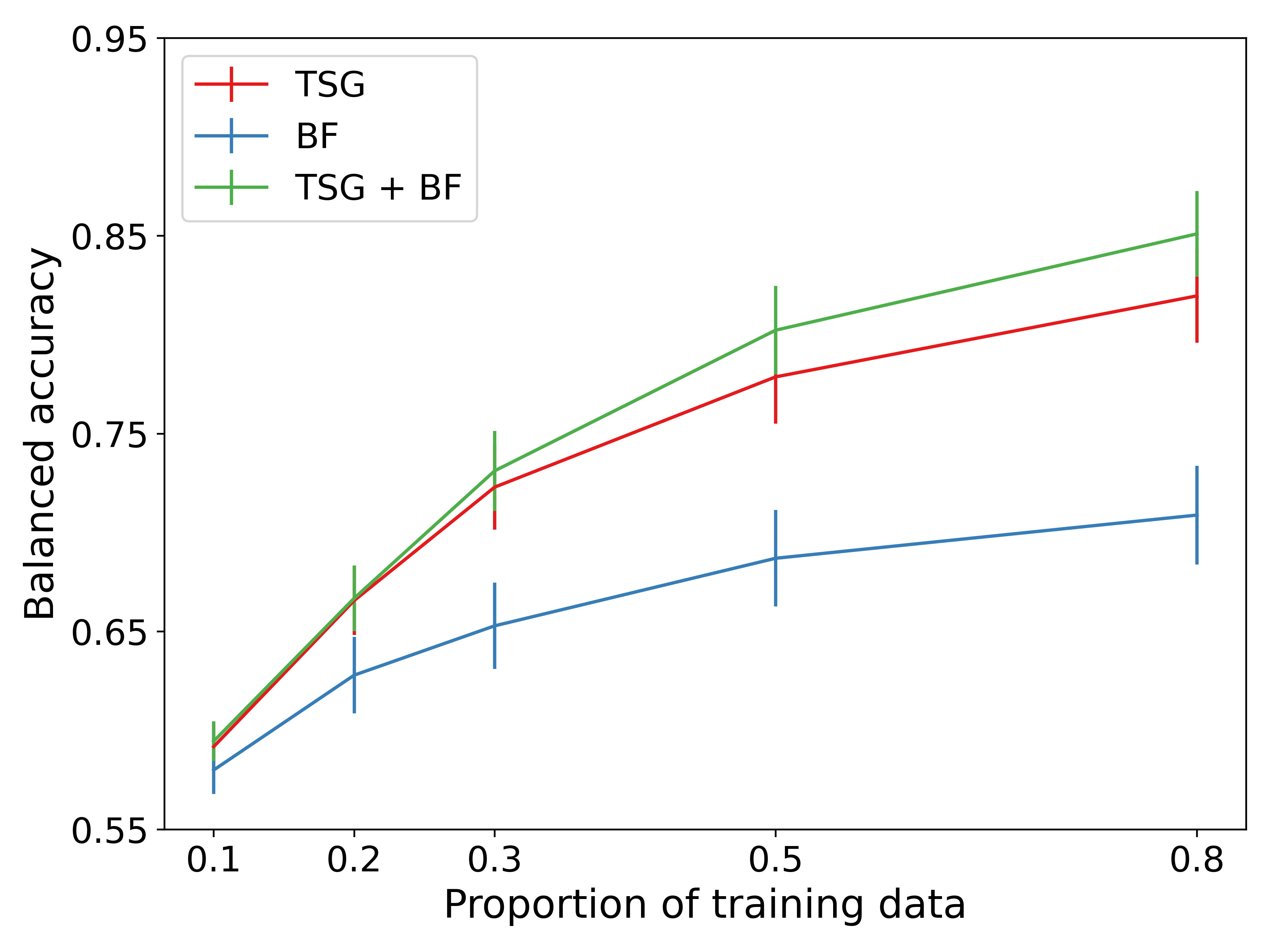}  
    \caption{Mental Math.}
    \end{subfigure}
    \begin{subfigure}{0.49\textwidth}
    \includegraphics[width=\linewidth]{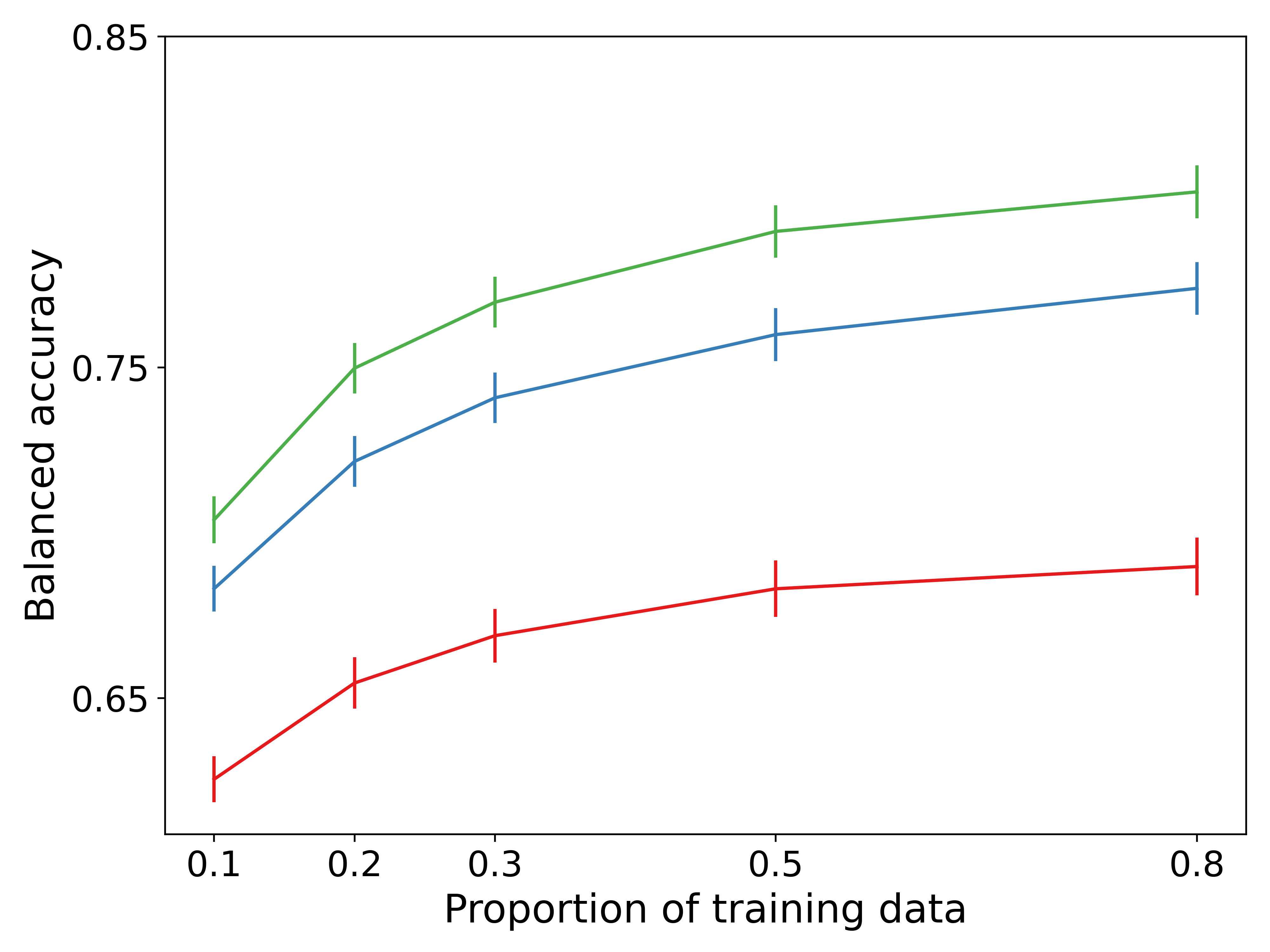}  
          \caption{MATB-II.}
          \label{}
    \end{subfigure}
     \\
    \begin{subfigure}{0.49\textwidth}
    \includegraphics[width=\linewidth]{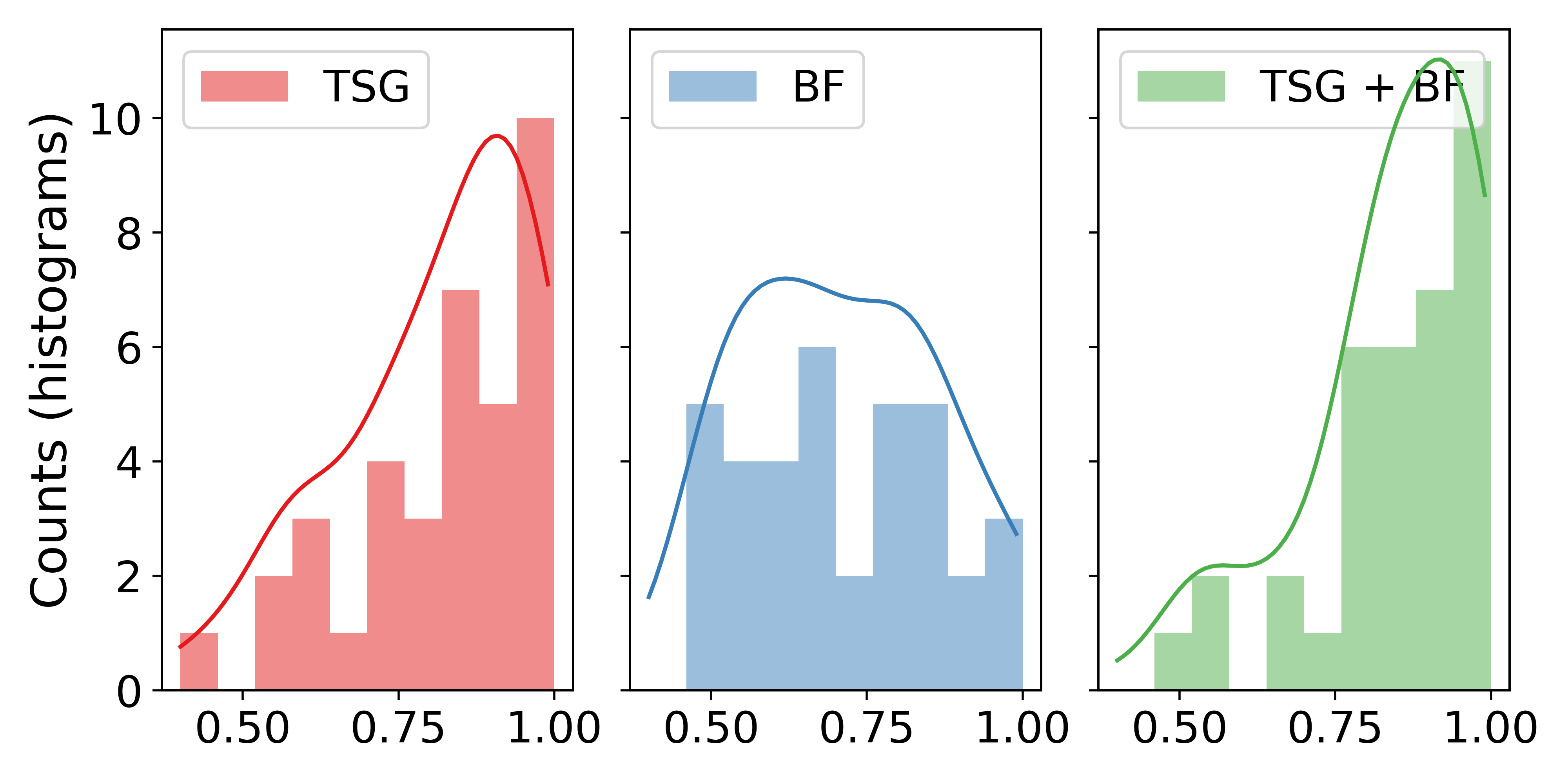}  
          \caption{Histogram and estimated densities of in-session subject balanced accuracies at p=0.8 for Mental Math.}
          \label{}
    \end{subfigure}
    \begin{subfigure}{0.49\textwidth}
    \includegraphics[width=\linewidth]{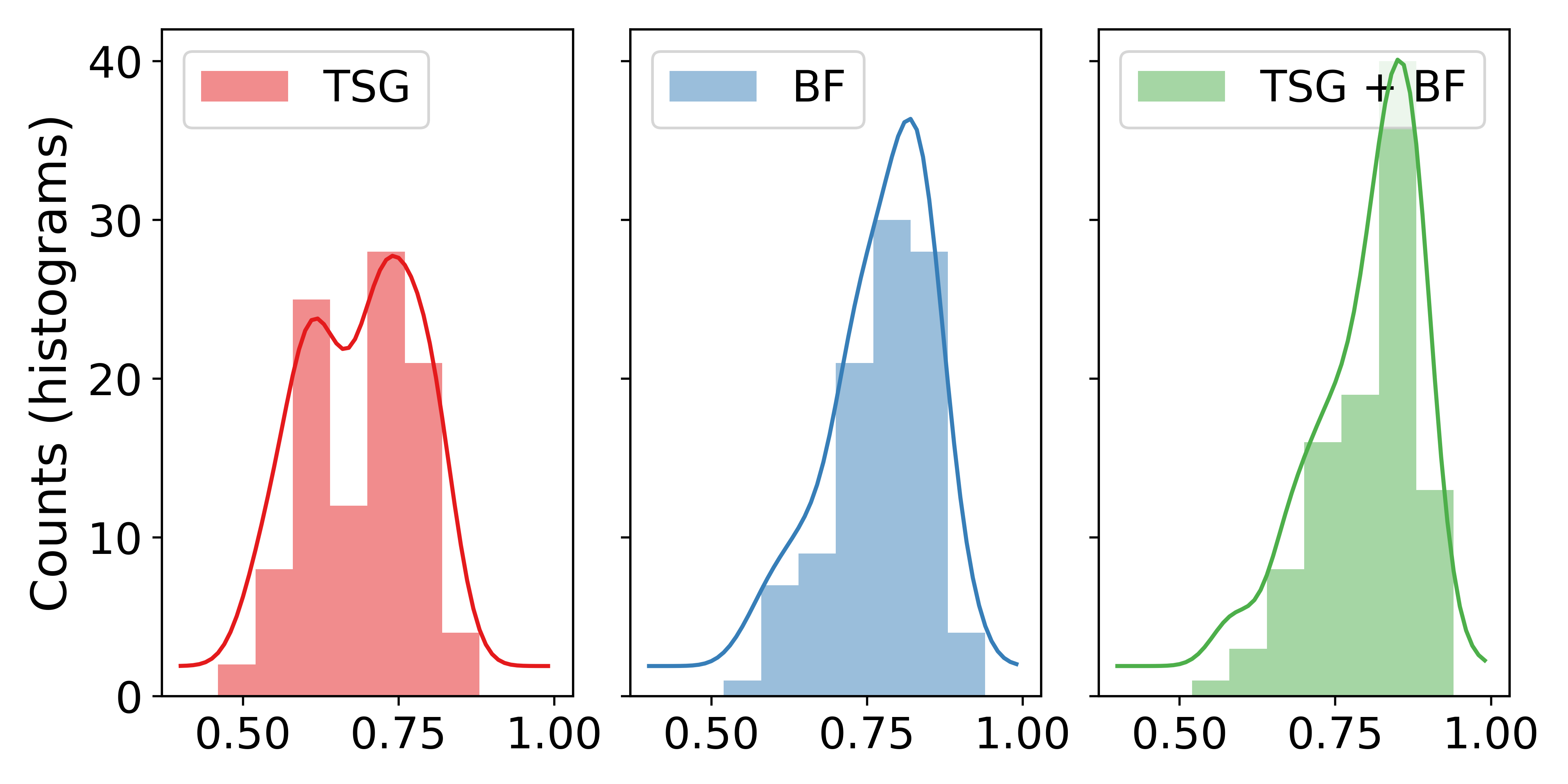}  
            \caption{Histogram and estimated densities of in-session subject balanced accuracies at p=0.8 for MATB-II.}
          \label{}
    \end{subfigure}
    \caption{In-session classification results.
    }
\label{fig:in-session}
\end{figure}

\subsection{Classification experiments}
\label{subsec:classification}
We study the three sets of features in in-session, non-constant querying, and transfer learning contexts derived from the two class classification problems \{stressed, not stressed\} and \{no \& low cognitive load, medium \& high cognitive load\}. The experiments attempt to mimic real-world use cases for passive BCIs: the in-session experiment will inform us on how much data we need to be performant for an average participant given no auxillary data (e.g., data from additional sessions, tasks or participants); the passive querying experiment gives us an idea of how often a system must query the participant for a label when passively recording EEG signal; and the transferability experiment will quantitatively (via balanced accuracy) measure the utility of pre-trained and lightly fine tuned models from other sessions and other subjects. In all, we think this set of experiments provides a useful suite of results to evaluate the considered feature sets and to inspire more nuanced experiments.

For all the experiments we use scikit- learn's \cite{scikit-learn} implementation of a random forest as the classifier after learning the appropriate set of features. The default hyperparameters (e.g., number of trees, maximum depth of each tree, impurity, etc.) from version 0.22 are used. We report the balanced accuracy for all experiments except for the sensor importance analysis for which we report standard accuracy. Error bars correspond to the standard error of the average accuracy across subjects. We provide a summary table in Table \ref{tab:summary} to more easily compare results across experiments.



\subsubsection{In-session classification}
\label{subsubsec:in-session}

The first experiment we consider is a standard classification experiment: we split the in-session data into a training set and a testing set, learn the feature map and classifiers using only the training set, and evaluate the classifier using the testing set. The top panels of Figure \ref{fig:in-session} show the average performance across subjects of the three sets of features for varying amounts of data available for training. For the Mental Math study (left), we see the graph-based features outperform the band-based features and their concatenation equal to the best throughout. For the MATB-II study (right), we see the band-based features outperform the graph-based features and their concatenation outperform both for the entirety of the studied regime. The result that TSG > BF for Mental Math but BF > TSG for MATB-II combined with the result that their concatenation is always preferred to either one suggests that the predictive information contained in the TSG and BF features is somewhat complementary -- there exists information in one set that is not present in the other. 

The bottom panels of Figure \ref{fig:in-session} show the histograms of the accuracies for each participant when the feature maps and classifiers have access to 80\% of the data for training. The left tail of the histograms shows that the performance is highly dependent on the subject. For example, in the Mental Math study (left) we see a non-trivial number of participants at chance level for each method even with 80\% of the data available for training. On the other hand, we also see numerous participants with a balanced accuracy well above 90\%. This effect is less exaggerated for the MATB-II data (right) but is still present for the band-based features and the concatenated features.

While not particularly informative of how real passive BCI predictive systems work, the performance of the systems in this experiment is somewhat of an upper bound for what we can expect from a real-time system when there is no additional data from other sessions or subjects. In particular, in this experiment the standard assumption of independence between the training and testing sets is likely violated due to correlations between adjacent windows. This can inflate classification performance \cite{hand2006classifier}.

The lines shown in Figure \ref{fig:in-session} are estimated averages over all of the subjects. For each subject we estimated their corresponding average using 45 random train-test splits for reach amount of training data. The histograms on the bottom row of Figure \ref{fig:in-session} show the distribution of these averages when the system has access to 80\% of the data. 

\begin{figure}[t]
    \captionsetup[subfigure]{justification=centering} 
    \begin{subfigure}{0.49\textwidth}
        \includegraphics[width=\linewidth]{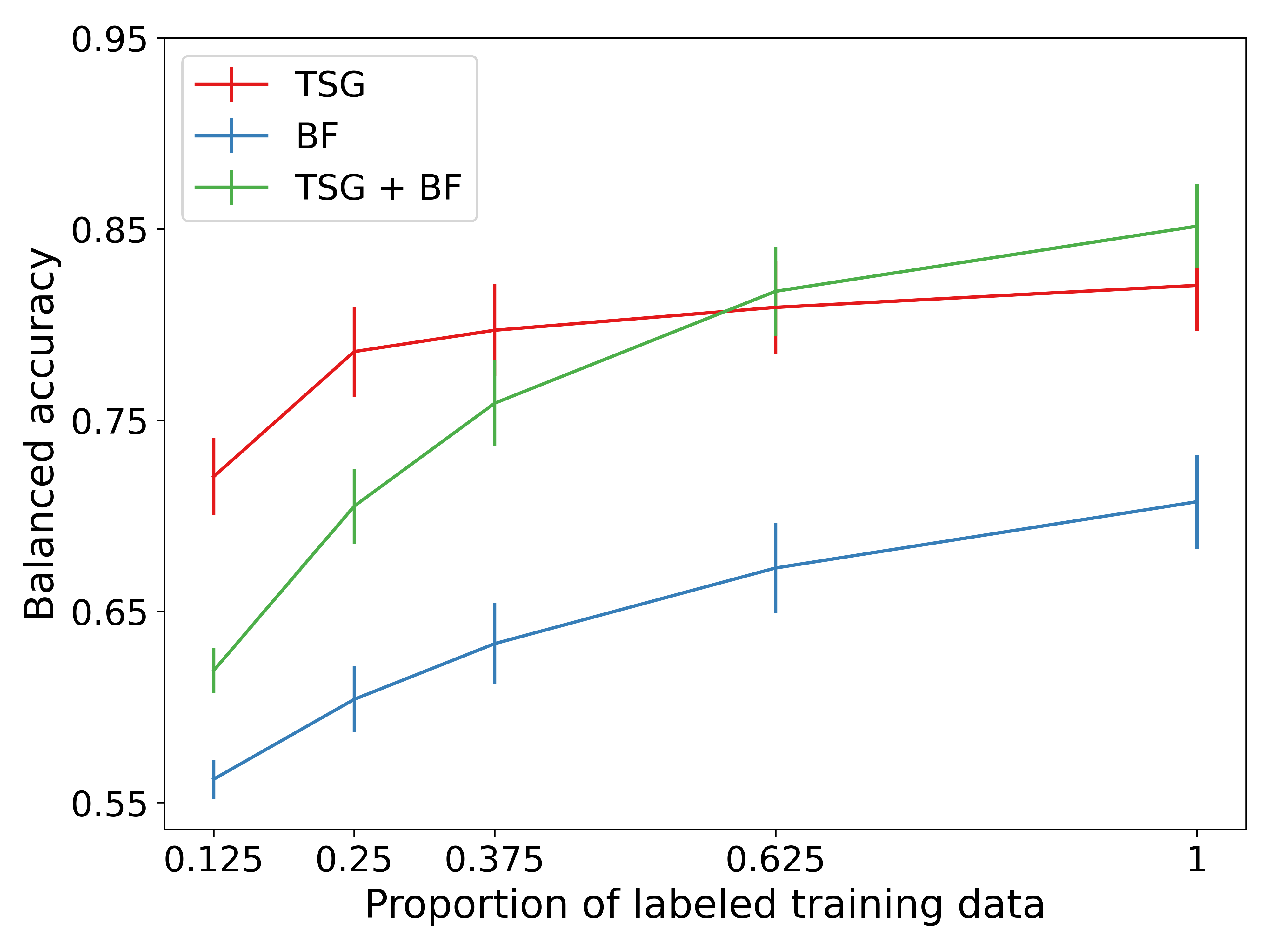}  
          \caption{Mental Math}
    \end{subfigure}
    \begin{subfigure}{0.49\textwidth}
    \includegraphics[width=\linewidth]{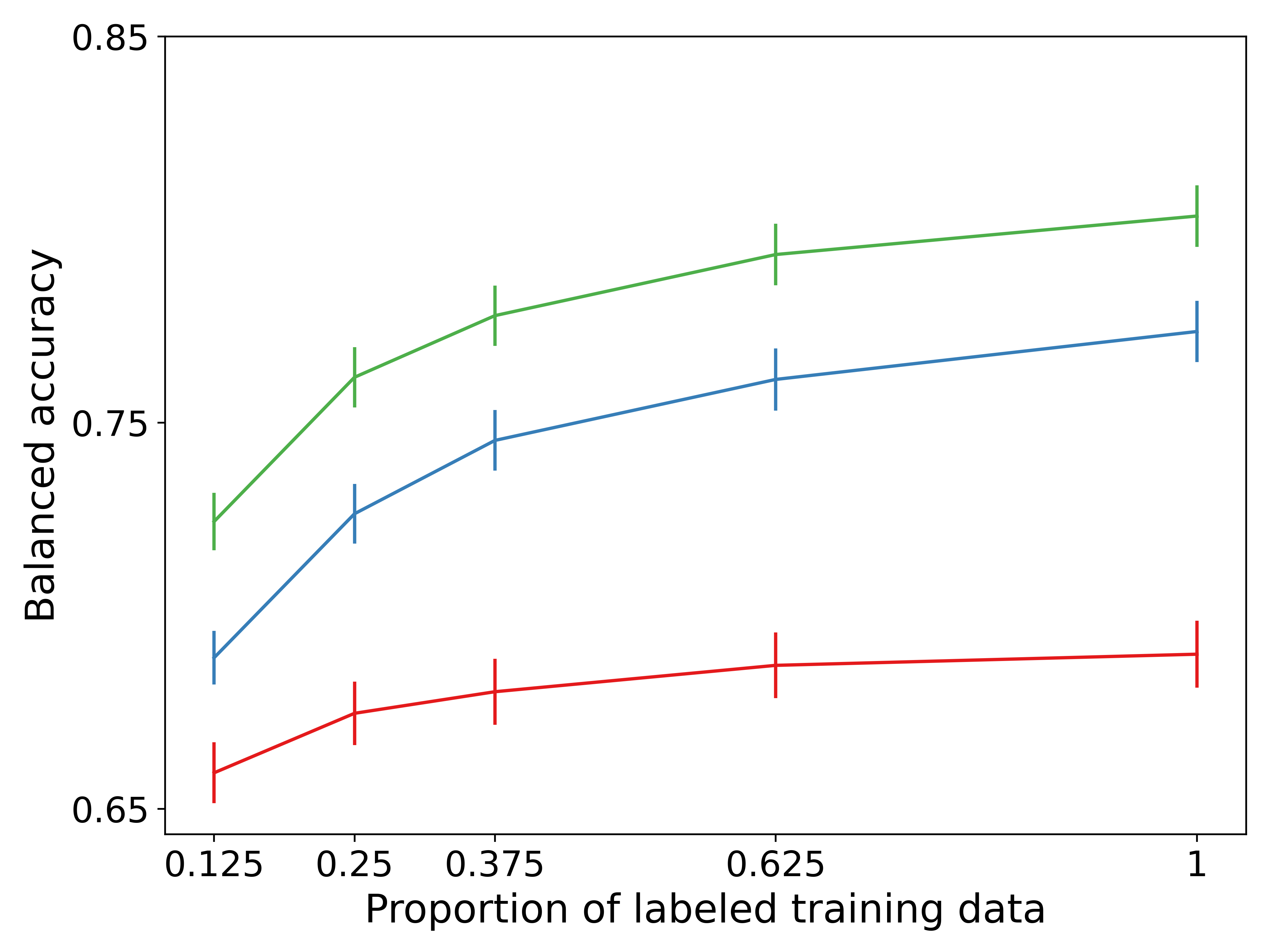}  
          \caption{MATB-II}
          \label{}
    \end{subfigure}
    \caption{Non-constant querying (semi-supervised) classification results.
    }
\label{fig:non-constant-querying}
\end{figure}
\subsubsection{Non-constant querying}
The next experiment we consider is inspired by the scenario where a participant is wearing an EEG headset throughout the day and the system has a the capability to query the user for a self-reported \{stress, not stressed\} or \{no \& low load, medium \& high load\} label for the most recent time windows. In these scenarios there is often a non-trivial amount of in-distribution but unlabeled data that can be leveraged to improve the representations ultimately used for inference. 

For this experiment, as in the in-session experiment, we randomly split the data into a training set (here always 80\%) and a testing set (20\%). We then further split the training set into an unlabeled set and labeled set and evaluate systems for varying ratios of labeled data to training data. This means that when the ratio is 0.125 the feature extraction methods have access to 80\% of all of the data and the classifier has access to only 10\% of the 80\%. When the ratio is equal to 1 then 80\% of all of the data is used for feature extraction and to train the classifier. Note that here we do not need to change our feature extraction techniques to accommodate unlabeled data because both methods are entirely unsupervised. More involved methods may require treating the labeled and unlabeled data differently \cite{zhou2014semi}.

Figure \ref{fig:non-constant-querying} shows the average performance across subjects for the three semi-supervised methods. Of note is the performance of TSG on the Mental Math study in the regime where the majority of the data is unlabeled relative to the same amount of labeled data for the in-session experiment (the x-tick 0.125 in Figure \ref{fig:non-constant-querying} corresponds to the x-tick 0.1 in Figure \ref{fig:in-session}). The boost is less impressive in the MATB-II study but still present. 

The band-based features are not meaningfully improved with the availability of a set of unlabeled data which likely implies that the principal components of the data are relatively stable across windows. The gain in performance of the concatenated features is somewhere in between the two composite features. We see a small (1-3\%) improvement over the corresponding in-session result for the first half of the regime.

The boosts in performance for the graph-based features and the concatenation of the features implies that the data used between queries in real-world systems can be used to improve predictive performance. This means that fewer queries can be made and users will be less likely to tire from self-reporting labels when leveraging the collected but unlabeled data.

This experiment likewise suffers from the effect of adjacent windows being in different splits of the data. The averages in Figure \ref{fig:non-constant-querying} are estimated using 45 random train-test splits. 

\begin{figure}[!t]
    \captionsetup[subfigure]{justification=centering}\centering
    \begin{subfigure}{0.49\textwidth}
    \includegraphics[width=\linewidth]{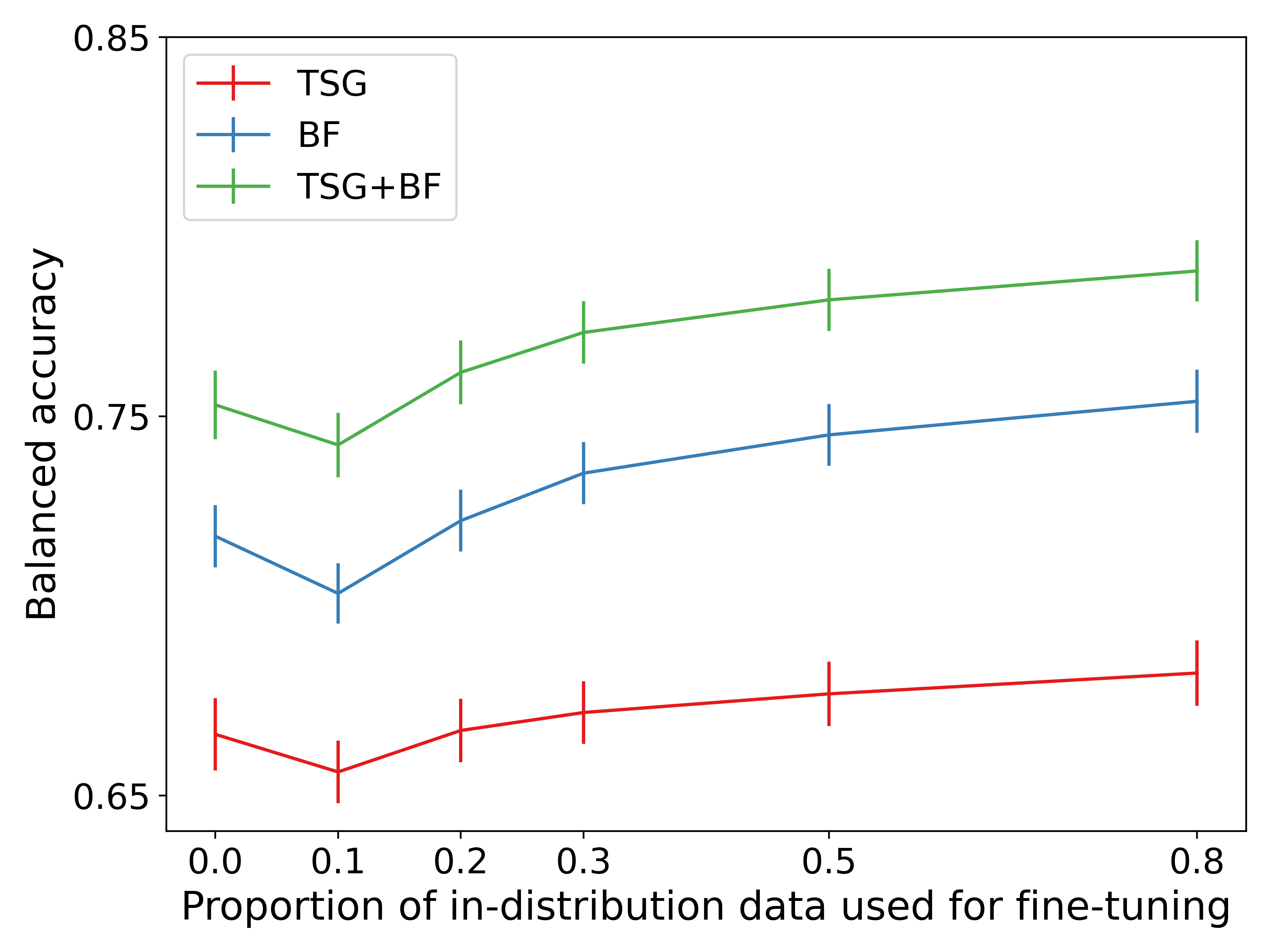}  
          \caption{MATB-II (across session)}
          \label{}
    \end{subfigure}
    \\
    \begin{subfigure}{0.49\textwidth}
        \includegraphics[width=\linewidth]{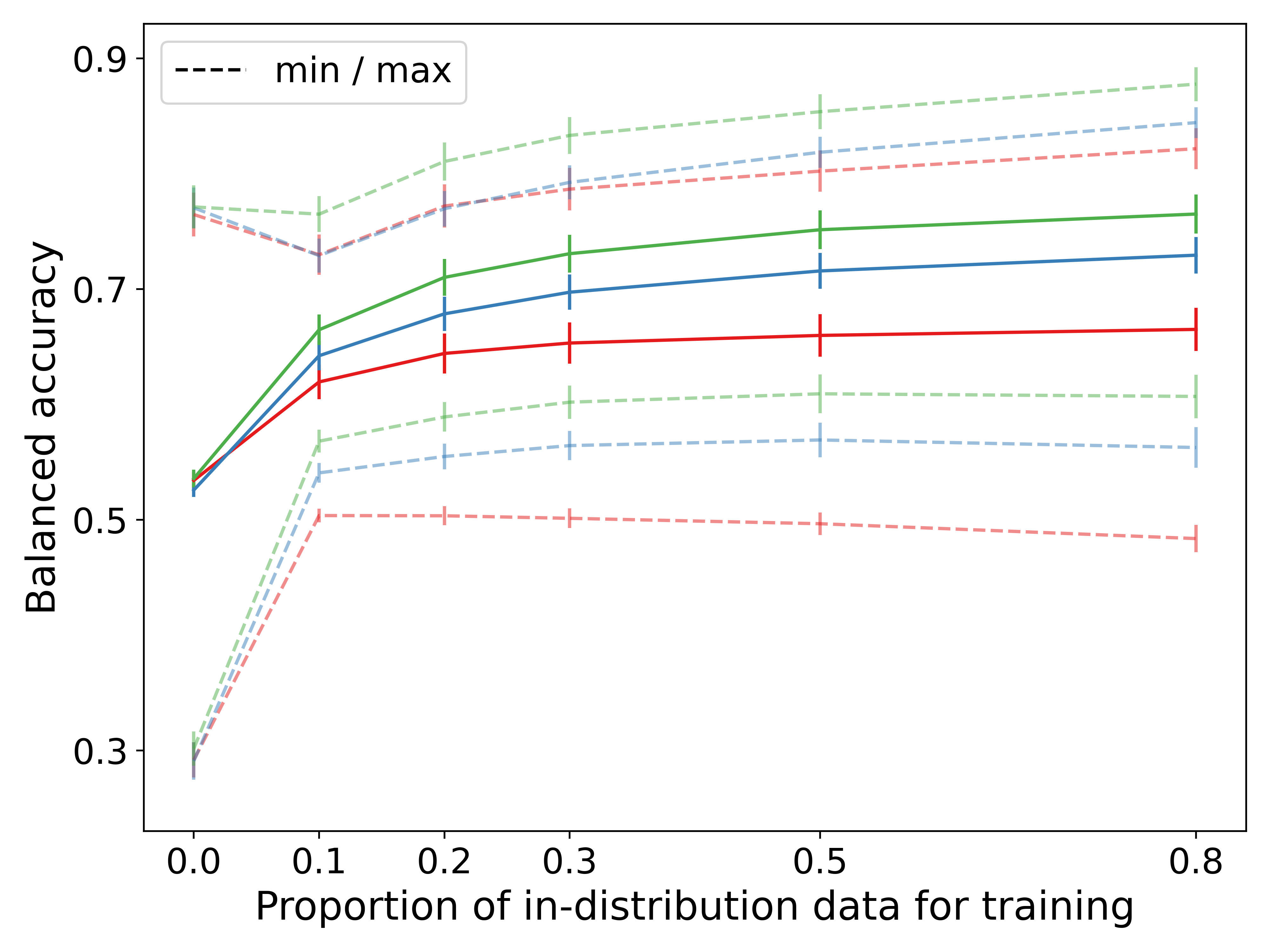} 
          \caption{Mental Math (across subject)}
    \end{subfigure}
    \begin{subfigure}{0.49\textwidth}
    \includegraphics[width=\linewidth]{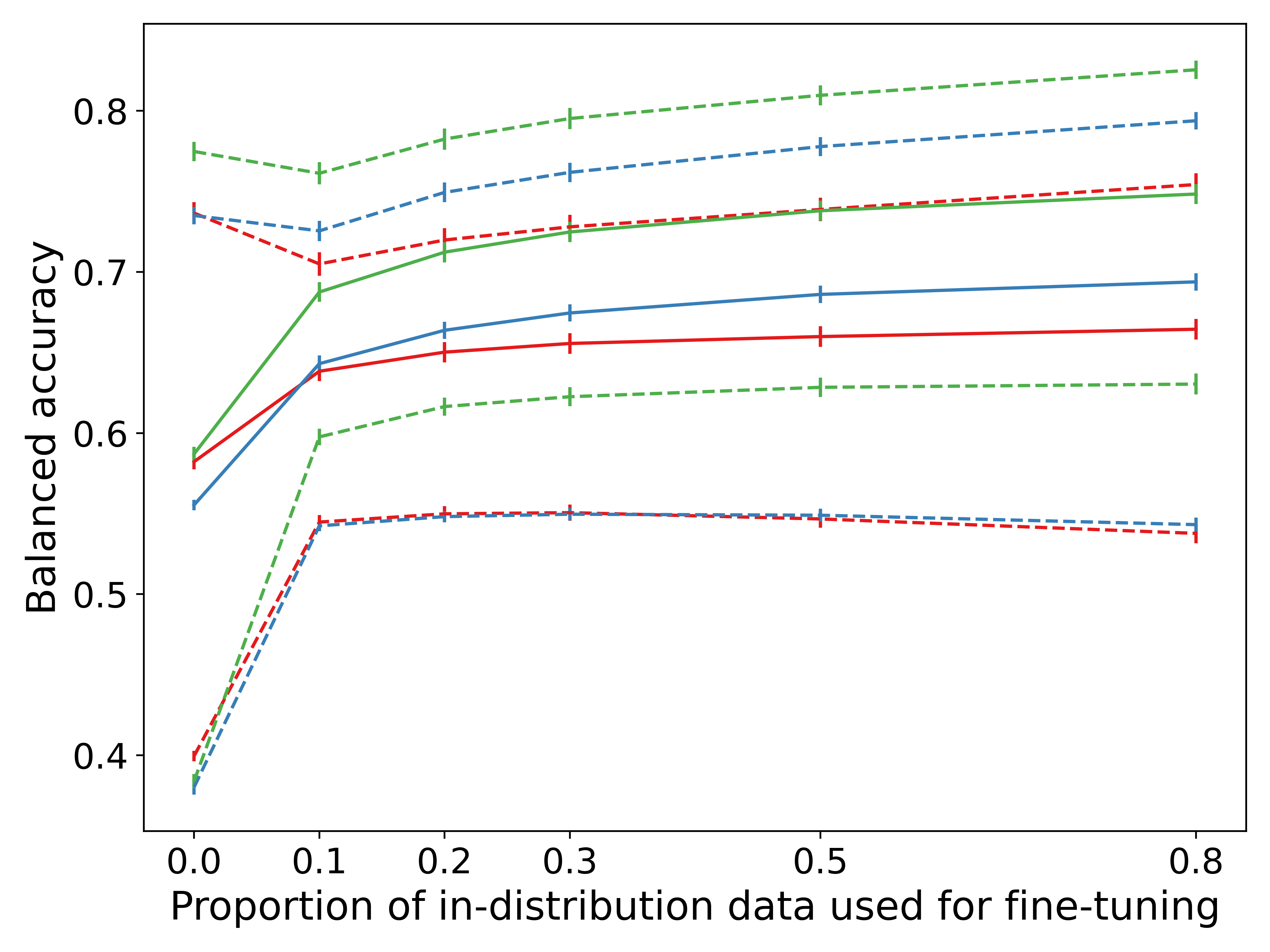}  
     \caption{MATB-II (across subject)}
          \label{}
    \end{subfigure}
    \caption{Transfer results.
    }
\label{fig:transfer}
\end{figure}
\subsubsection{Transferability}

The last classification experiment we consider investigates the transferability of feature maps and classifiers from one session to another. Successful zero-shot or few-shot transfer for all participants would enable rapid deployment of real-time prediction systems for a variety of tasks. Unfortunately, as with many biosignals, the inter-session and inter-subject variability of EEG signals means that pre-trained models are unsuited for prediction related to a new participant \cite{wei2021inter}. Our transfer experiments are designed to help understand the limitations and possibilities of single-session transfer in the context of mental state prediction.

We consider two transfer paradigms: zero-shot learning \cite{xian2017zero} and classifier fine-tuning. For zero-shot learning we simply take the feature maps and classifiers learned from another session and apply it to a new session. These approaches are rigid and only work well when the distribution shift between sessions is small. For classifier fine-tuning we use the feature map from another session and transform the data from the new session into the appropriate feature space. We then use data from the new session to update the classifier trained on the other session. Since we are using random forests, this simply means taking the existing structure of the trees and updating the posteriors in each of the leaf nodes for each tree with new labeled data. The posteriors of empty leaf nodes are set to 0.5 for simplicity. In the results of Figure \ref{fig:transfer} the zero-shot performance corresponds to the x-tick 0.0 and the rest of the figure corresponds to classifier fine-tuning.

The top panel of Figure \ref{fig:transfer} shows the zero-shot and classifier fine tuning results when transferring from one MATB session to another for a given subject. We again plot the average across subjects as in the previous two experiments.

In the MATB experimental design the two sessions are close in time and thus the distribution shift is relatively small as is evident by the good zero-shot performance across methods. For TSG the zero-shot is effectively as good as the fine tuning for the majority of the regime. For all methods the zero-shot performance and is approximately as good as having access to about 30\% of the in-session data. Recall that in the MATB experiment each session is approximately 50 minutes, meaning the zero-shot approach across sessions can save approximately 50 (0.3) = 15 minutes of calibration time and data collection.

The performance across methods degrades non-trivially when going from zero-shot to parts of the regime where only a small amount of training data is available. This dip in performance is again indicative of the proximity of the distributions between sessions and can be smoothed out by continual or lifelong learning techniques \cite{vogelstein2022representation, thrun1995learning, geisa2022theory}. Unfortunately there is no concept of transferring across sessions for the Mental Math study and so there is no corresponding figure.

The bottom row of Figure \ref{fig:transfer} contains the cross-subject transfer results for both Mental Math (left) and MATB-II (right). For each subject there are $ n_{subjects} - 1 = 35$ corresponding accuracies for the Mental Math data and $ n_{subjects} \cdot n_{sessions} - 2 =98 $ corresponding accuracies for the MATB-II data. From these accuracies we record the average, minimum accuracy, and maximum accuracy when transferring to each subject. The average across subjects of these three statistics for both datasets are shown in the respective figures.

The range between the averages of the minimums and maximums in both datasets is particularly interesting -- if you attempt to use zero-shot or few-shot classification from the wrong subject the performance is severely hampered; on the other hand, if you can correctly identify which feature map and classifier is ``best" you can get a serious improvement gain over the average performance. Selecting which model or set of models to transfer from is an active area of research \cite{helm2020partitionbased} and successful techniques will likely be of unusual importance in biosensing settings to keep on-device models as light, private, and effective as possible.  

The dip present in the cross-session figure is not present in the average cross-subject transfer curves (it is present in the ``max" curve) and is indicative of the average out of subject distribution being sufficiently far from the distribution we are attempting to transfer to.

\subsubsection{Summary table} Table \ref{tab:summary} summarizes the results across the three experiments for easier comparison across the different settings. Note that for non-constant querying the x-ticks in Figure \ref{fig:non-constant-querying} are multiplied by 0.8 to get to the appropriate amount of labeled training data (e.g., 0.125 (0.8) = 0.1).

\begin{table}[]
    \centering
    \begin{tabular}{||c||c|c|c|c|c|c||}
     \hline\hline
         & MM (p=0.0) & MM (p=0.1) & MM (p=0.8) & MATB (p=0.0) & MATB (p=0.1) & MATB (p=0.8)
         \\
         \hline \hline
         TSG-1 & n/a & 59.2 (1.2) & 82.0 (2.4) & n/a & 62.6 (1.2)& 69.0 (1.5)
         \\
         TSG-2 & n/a & 72.1 (2.0) & 82.1 (2.4) & n/a & 65.9 (1.3) & 69.0 (1.4)
         \\
         TSG-3 & n/a & n/a & n/a & 66.6 (0.1) & 65.6 (0.8) & 68.2 (0.8)
         \\
         TSG-4 & 53.4 (0.7) & 62.0 (1.5) & 66.5 (1.8) & 58.2 (0.4) & 63.2 (0.6) & 65.9 (0.6)
         \\
         \hline
         BF-1 & n/a & 58.0 (1.2) & 70.9 (2.5) & n/a & 68.3 (1.2)  & 77.4 (1.3)
         \\
         BF-2 & n/a & 56.2 (1.0) & 70.7 (2.5) & n/a & 68.9 (1.2) & 77.4 (1.3)
         \\
         BF-3 & n/a & n/a & n/a & 71.8 (0.8) & 70.3 (0.7) & 75.3 (0.8)
         \\
         BF-4 & 52.8 (0.5) & 64.2 (1.2) & 72.9 (1.5) & 55.7 (0.2) & 63.6 (0.5) & 68.8 (0.5)
         \\
         \hline
         TSG+BF-1 & n/a & 59.5 (1.0) & 85.1 (2.2) & n/a & 70.4 (1.2) & 80.3 (1.3)
         \\
         TSG+BF-2 & n/a & 61.9 (1.2) & 85.2 (2.2) & n/a & 72.4 (1.2) & 80.4 (1.3)
         \\
         TSG+BF-3 & n/a & n/a & n/a & 75.3 (0.9) & 74.2 (0.8) & 78.8 (0.8)
         \\
         TSG+BF-4 & 53.6 (0.7) & 66.5 (1.3) & 76.5 (1.7) & 59.0 (0.4) & 68.1 (0.6) & 74.3 (0.7)
         \\
         \hline
         \hline
    \end{tabular}
    \caption{Summary table of the in-session, non-constant querying, and transfer results for the Mental Math (``MM") and MATB-II (``MATB") datasets. Each column corresponds to a different proportion ($ p \in \{0.0, 0.1, 0.8\} $) of in-distribution data available per the experiments described in the first three subsections of Section \ref{subsec:classification}. For each method \{TSG, BF, TSG+BF\} there are four variations (1 = in-session, 2 = non-constant querying, 3 = cross-session transfer, 4 = cross-subject transfer). Values are average balanced accuracies across subjects. Standard errors are given in parentheses.}
    \label{tab:summary}
\end{table}

In the majority of contexts explored above the combination of the the band-based and graph-based features out performed either on their own and, again, suggests that both should be used to capture as much predictive information as possible.

\begin{figure}
    \centering
    \begin{subfigure}{0.49\textwidth}
    \includegraphics[width=\linewidth]{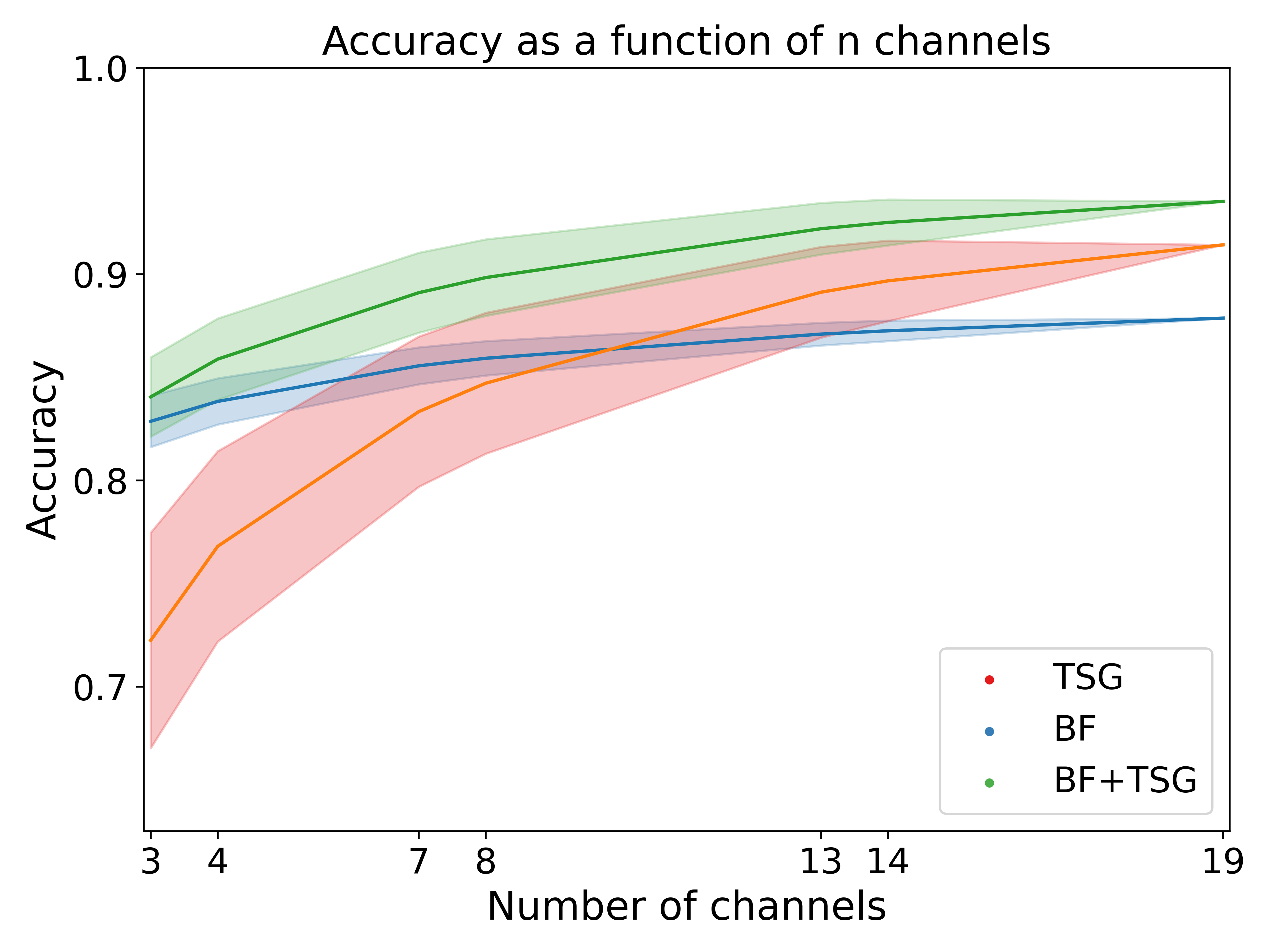}
    \caption{Combinatorial search over sensor configurations with corresponding accuracies. Shaded regions contain 90\% of the accuracies for a given number of subsets.}
    \label{subfig:sensor-selection}
    \end{subfigure} \\
    \centering
    \begin{subfigure}{\textwidth}
    \includegraphics[width=\linewidth]{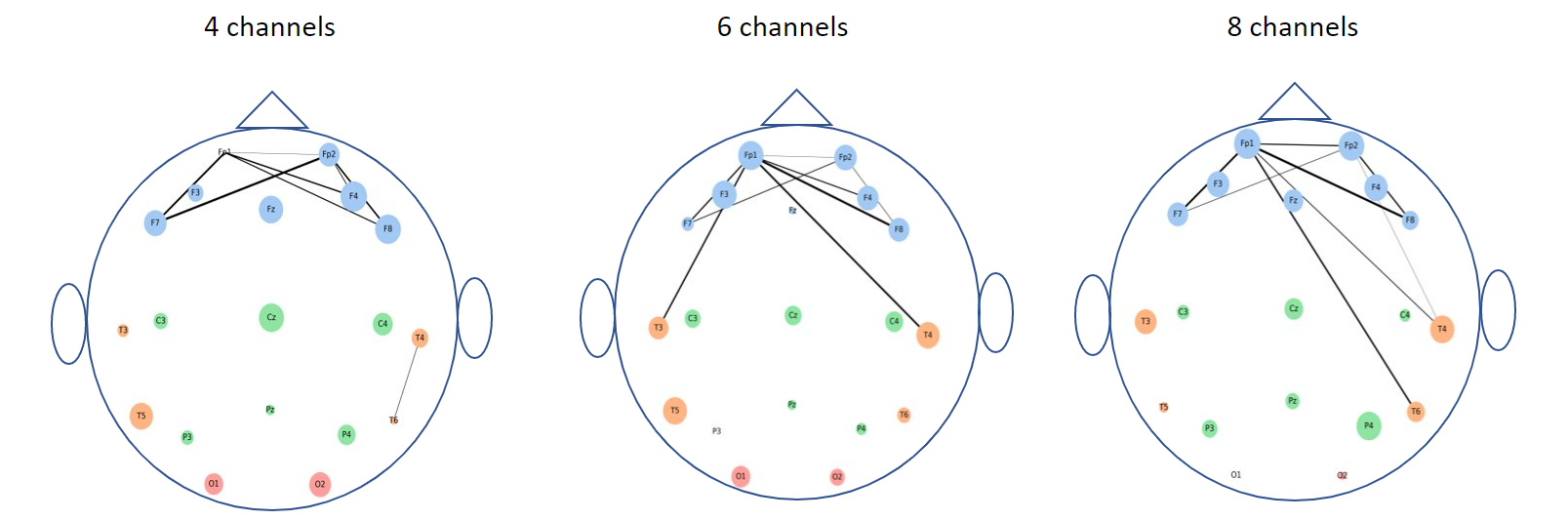}
    \caption{Channel importance (size of channel) for every channel and channel pair importance (line width of connection) for top subset of all pairs. \textbf{We do not show all edges so as not to crowd the figure.}}
    \label{subfig:channel-importance}
    \end{subfigure}
    \caption{Sensor and pair of sensors analysis. The top figure shows the distribution of averages for varying numbers of channels for TSG, BF, and TSG+BF. The bottom figure shows the importance of each sensor and pair of sensors: the size of the circles for each channel is proportional to its rank amongst channels; the line width of the edge between two channels is proportional to the pair's rank amongst pairs of channels; he color of the channel is indicative of the region of the head (frontal, temporal, central, occiptal).}
\end{figure}

\subsection{Sets of sensors analysis and channel and pairs of channels importance study.}

While the classification results of Section \ref{subsec:classification} are interesting in their own right, the sensor locations and device set ups used in the experimental procedures are relatively arbitrary and are not subjected to the same design constraints as a real-world system. Indeed, commercially viable EEG hardware needs to be performant with a relatively minimal head coverage and number of sensors. To understand the effect of removing sensors on classification performance, we re-ran the in-session classification experiment from Section \ref{subsubsec:in-session} with all subsets of \{3,4,5,6,7,8,13,14,18,19\} channels and recorded their accuracy on the Mental Math data. For each subset we estimated the average accuracy using 30 random train-test splits with 80\% of the data used for training for each subject.

Figure \ref{subfig:sensor-selection} shows the average accuracy\footnote{We performed this analysis before the classification experiments of Section \ref{subsec:classification} and, importantly, before we began considering balanced accuracy instead of standard accuracy. Since the analysis is  combinatorial and thus computationally expensive we did not re-run the analysis. While the quantitative results change with the switch to standard accuracy, the qualitative results still hold.} across subsets as a function of the number of channels in the subset. Shaded regions indicate the 90\% interval around the mean subset accuracy. The subsets in the right tail of the distribution of subsets are notably close in performance to the full suite of channels for the concatenated features with access to only 3 or 4 channels. This implies that a performant system is possible with only a small selection of channels. 

We can understand which channels and pairs of channels are driving the high accuracy with such a small subset by estimating their importance. For each channel we first calculate the average accuracy on the stress prediction task for subsets that it is an element of. We then rank the channels based on their average accuracy. We can do the same for pairs of channels. Figure \ref{subfig:channel-importance} is a pictorial description of these rankings for subsets of size 4, 6, and 8: the size of the circles for each channel is proportional to its rank amongst channels and the line width of the edge between two channels is proportional to the pair's rank amongst pairs of channels. The color of the channel is indicative of the region of the head (frontal, temporal, central, occiptal). 

The channel importance and pair of channel importance for each 4, 6, and 8 channel subsets are well aligned with conventional stress-related neuroscience wisdom. In particular, the average importance of individual channels in the frontal region (blue channels) is noticeably larger than the average importance of the other regions \cite{Yuen14075}. Further, there are a considerable amount of heavily weighted edges between frontal channels on one side of the brain to frontal or temporal channels on the other side of the brain \cite{10.3389/fnbeh.2018.00166}. Hence, we think that including the pairwise correlation matrix as the basis for a set of features is neuroscientifically reasonable.

\section{Discussion}
\label{sec:discussion}
In this note we proposed a set features, TSG, that are jointly learned from the correlation matrices of different windows. We showed in multiple classification paradigms that the TSG features can be used to significantly improve performance over canonical band-based features, BF, in both stress and cognitive load contexts. We noted that neither the TSG features nor the BF features dominate the other across the different experiments and thus argued that these two sets of features are ``complementary", or that one set of features contains predictive information that the other does not. Given these results we think that graph-based features will be useful in pushing the community closer to performant non-invasive BCIs.

We want to emphasize that the proposed set of features are but one instance of a class of features based on the implicit graph structure on the channels. We plan to investigate other natural similarities on the channels beyond just Pearson correlation, e.g., mutual information and conditional entropy, and think that these more nuanced statistical dependence measures could capture more meaningful class-conditional information.

Further, the similarity that we use to induce a graph on the collection of statistical-dependence matrices is Frobenious norm based and as simple as posssible. It is well known in the network statistics community that the Frobenious norm between matrices is too crude of a metric to uncover particular asymptotic behaviors of the graph \cite{kriege2020survey, JMLR:v18:17-448}. Indeed, often dimensionality reduction techniques are used on the collection of graphs before calculating the distance between the objects. We think that including a multi-graph embedding step such as the omnibus embedding \cite{levin2017central} or the COSIE framework \cite{arroyo2020inference} could non-trivially improve performance over the results presented here. Similarly, we think applying different kernels on the networks, such as the Gaussian kernel, could be a fruitful direction.


\section*{Acknowledgements}
We would like to thank Ben Cutler, Steven Dong, Amber Hoak, Tzyy-Ping Jung, Ben Pedigo, Siddharth Siddharth, and Christopher White for helpful comments and suggestions throughout our investigations.

\bibliographystyle{unsrt}  
\bibliography{references}  






\end{document}